\documentclass[journal]{IEEEtran}   
\usepackage[utf8]{inputenc}
\usepackage[T1]{fontenc}
\usepackage{graphicx}
\usepackage{subcaption}  
\usepackage{caption}
\usepackage{adjustbox}   
\usepackage{epstopdf}
\usepackage{hyperref}
\usepackage{placeins}  
\usepackage{amsmath,amssymb,amsthm}
\usepackage{mathtools}
\usepackage{bm}
\usepackage{graphicx}
\usepackage{subcaption}
\usepackage{caption}
\usepackage{float}
\usepackage{booktabs}
\usepackage{xcolor}
\usepackage{enumitem}
\usepackage{verbatim}
\usepackage[mathscr]{euscript}
\usepackage{epstopdf}
\usepackage{hyperref}

\DeclareMathOperator{\tr}{tr}

\DeclareMathOperator{\Det}{det}

\newcommand{\HH}{\mathsf{H}}
\newcommand{\EE}{\mathbb{E}}

\newcommand{\CN}{\mathcal{CN}}
\newcommand{\IG}{\mathcal{IG}}

\newcommand{\condi}{\,|\,}

\begin{document}

\title{Robust Expectation–Maximization for Covariance Estimation in SIRV Models with Missing Data: Application to InSAR Time Series}

\author{
    M. Cherifi\thanks{M. Cherifi is with the Laboratoire Traitement du Signal,
    École Militaire Polytechnique (EMP), Bordj El Bahri, Algiers, Algeria
    (e-mail: cherifi.meddd@gmail.com).},
    M. N. El Korso\thanks{M. N. El Korso is with CentraleSupélec,
    Université Paris-Saclay, Gif-sur-Yvette, France
    (e-mail: mohammed.nabil.el-korso@centralesupelec.fr). He is a Member of IEEE.},
    A. Hippert-Ferrer\thanks{A. Hippert-Ferrer is with LASTIG, IGN Géodata Paris,
    Université Gustave Eiffel, Champs-sur-Marne, France
    (e-mail: alexandre.hippert-ferrer@ign.fr).},
    and Y. Yan\thanks{Y. Yan is with LISTIC, Polytech Annecy-Chambéry,
    Université Savoie Mont Blanc, Annecy, France
    (e-mail: yajing.yan@univ-smb.fr).}
}

\markboth{IEEE Transactions on Geoscience and Remote Sensing}%
{Cherifi \MakeLowercase{\textit{et al.}}: Robust EM for SIRV Covariance Estimation}

\maketitle

\begin{abstract}
This paper presents a comprehensive Expectation-Maximization (EM) framework for robust covariance estimation in Scale-Invariant Random Vector (SIRV) models with missing data under ignorable missingness mechanisms. {We focus on a practically important subclass of SIRVs by assuming conjugate inverse-gamma priors on the scale variables, which leads to a complex multivariate Student-t observation model.} We derive closed-form solutions for the E-step {for this subclass} by leveraging the conjugacy {property}, enabling efficient estimation while maintaining theoretical coherence. The proposed algorithm incorporates numerical robustness techniques including computation reuse for common observation patterns, regularized matrix inversions, and explicit enforcement of Hermitian positive semi-definite structure. Validation on synthetic data and Sentinel-1 interferograms demonstrates effective missing value retrieval and denoising performance even with limited data and correlated gaps.
\end{abstract}


\section{Introduction}

Reliable reconstruction of covariance matrices from incomplete observations is a central problem in signal processing and remote sensing. In contexts such as synthetic aperture radar (SAR) imaging or wireless communications, measurements often exhibit heavy-tailed distributions and local energy fluctuations (speckle), for which classical Gaussian models are inadequate \cite{conte2002recursive, chitour2008exact}. Scale-invariant random vector (SIRV) models provide a natural representation of these phenomena by separating a complex Gaussian directional component from a positive random scale factor that accounts for texture and amplitude variability \cite{delmas2024elliptically,pascal2008covariance,tyler1987distribution}. Robust structured covariance estimation under such elliptical models has been successfully applied to detection and estimation problems in radar imaging \cite{meriaux2020matched,abdallah2019detection}, where the same non-Gaussian, heavy-tailed behavior that motivates robust detectors also motivates the robust estimation framework developed in this paper.

Among the SAR applications affected by these non-Gaussian phenomena, surface displacement time series derived from interferometry (InSAR) are particularly challenging, as they commonly contain spatial and temporal gaps caused by decorrelation, algorithmic limitations, or surface changes. Restoring these time series while preserving their dynamic components (trends, oscillations) as well as the spatial and temporal correlation of noise is a recurring challenge in geoscience and remote sensing \cite{kondrashov2006spatio, gerber2018predicting}. Various methods exploit spatio-temporal structure to fill in missing data, for instance through factorial models, modal decompositions, or spatio-temporal kriging \cite{beckers2003eof, hippert2020eof,hippert2022robust,little2002statistical}. However, these imputation methods are generally built on Gaussian assumptions: when observations display strongly non-Gaussian behavior or contain significant outliers, estimators based on the empirical covariance may be biased. Adopting a statistical framework that explicitly accounts for non-Gaussianity, in the spirit of the robust SIRV-based estimators discussed above, can thus improve estimator robustness in these situations \cite{chitour2008exact, chen2018robust}.

Although robust covariance estimation and missing data imputation have been extensively studied separately, their intersection remains relatively unexplored. A recent comprehensive survey \cite{hippert2025missing} provides a unified view of missing data models, methods, and modern approaches, motivating the joint treatment pursued in this work. Related robust inference frameworks with incomplete data have also been explored in other statistical models, such as logistic regression \cite{cherifi2025robust}. Standard Expectation-Maximization (EM) algorithms for Gaussian data with missing values lack robustness to outliers and non-Gaussianity \cite{dempster1977maximum}, while robust estimators based on SIRVs generally assume complete data \cite{ollila2012complex,frahm2010generalization,sun2014regularized}. This gap is particularly critical in applications such as SAR interferometry, where data are intrinsically non-Gaussian and frequently incomplete.

This article addresses this gap by developing a unified EM framework tailored to a practically important subclass of SIRV models with missing data under ignorable missingness mechanisms, namely Missing Completely At Random (MCAR) and Missing At Random (MAR). Specifically, by placing a conjugate inverse-gamma prior on the scale variables, the resulting observation model reduces to a complex multivariate Student-t distribution, for which closed-form E-step and M-step updates can be derived. The main contribution of this work is the complete derivation of this EM algorithm for SIRV covariance estimation with missing data, providing closed-form analytical expressions for both steps by exploiting the conjugacy between the inverse-gamma prior and the complex Gaussian likelihood. Our approach preserves the full SIRV structure during imputation, thereby avoiding the trade-offs between robustness and efficiency inherent to Gaussian EM methods or ad hoc imputation procedures. Comprehensive experimental validation on synthetic data and real Sentinel-1 interferograms demonstrates the method's effectiveness in estimating covariance matrices and imputing missing values under various missingness regimes and non-Gaussian characteristics.

We validate the method on synthetic datasets designed to mimic the key properties of interferograms and on a Sentinel-1 A/B time series. The experiments demonstrate that the EM--SIRV approach produces faithful imputations and yields covariance estimators more resilient to heavy tails and outliers than procedures based solely on the empirical covariance. The proposed framework provides a solid foundation for applications in radar imaging, remote sensing, and wireless communications, where non-Gaussian distributions and missing data are prevalent.

The remainder of this article is organized as follows. Section~2 presents the full derivation of the EM algorithm for SIRV models with missing data. Section~3 details the experimental validation on synthetic and real SAR data. Finally, Section~4 concludes with a discussion and perspectives for future work.
\section{Robust EM Algorithm for SIRV Models with Missing Data}

We derive the Expectation-Maximization (EM) algorithm for a scaled complex Gaussian (SIRV) observation model in which each spatial sample \(i\) is associated with two latent quantities: a positive scalar \(\tau_i\) that models local power variability (speckle) and a subset of missing complex components \(\bm{x}_{i,\mathrm{m}}\) of the multivariate observation \(\bm{x}_i\in\mathbb{C}^l\). The exposition establishes the complete-data likelihood, applies the law of iterated expectations to obtain closed-form expressions for the E-step under an inverse-gamma prior on \(\tau_i\), and derives the M-step closed form for the covariance \(\bm{\Sigma}\). Emphasis is placed on mathematical rigor and on numerical robustness: choice of conditioning order, reuse of computations across common observation patterns, regularized inversions, and explicit enforcement of Hermitian positive semi-definiteness. Notation: \(l\) denotes the ambient dimension, \(n\) the patch size, and \(\bm{x}_i=[\bm{x}_{i,o}^\top,\bm{x}_{i,m}^\top]^\top\) the partition into observed and missing blocks.

\subsection{Model Formulation and Priors}
{We consider the general SIRV observation model:}
\begin{equation}\label{eq:scaled-model}
\bm{x}_i \condi \tau_i,\bm{\Sigma} \sim \CN\bigl(\bm{0}, \tau_i\bm{\Sigma}\bigr),
\end{equation}
{where $\bm{\Sigma}\in\mathbb{C}^{l\times l}$ is an unknown Hermitian positive semidefinite (PSD) covariance matrix, and $\tau_i > 0$ is an unobserved scalar random variable modeling the local power. The marginal distribution of $\bm{x}_i$ is obtained by integrating over $\tau_i$ and is generally non-Gaussian and heavy-tailed.

The complete-data likelihood, considering both the observed vector and the latent scale, is given (up to a constant) by:}
\begin{equation}\label{eq:log-complete}
\log p(\bm{x}_i,\tau_i \condi \bm{\Sigma})
= -l\log\tau_i - \log\Det{\bm{\Sigma}} - \frac{1}{\tau_i} \bm{x}_i^{\HH}\bm{\Sigma}^{-1}\bm{x}_i + \log p(\tau_i).
\end{equation}

{To obtain a tractable Expectation-Maximization (EM) algorithm with closed-form updates, we must specify a prior $p(\tau_i)$. A common and convenient choice that retains conjugacy with the complex Gaussian likelihood is the inverse-gamma (IG) distribution:}
\begin{equation}\label{eq:tau-prior}
\tau_i \sim \IG(\alpha_0,\beta_0),\qquad p(\tau_i)\propto\tau_i^{-(\alpha_0+1)}\exp(-\beta_0/\tau_i).
\end{equation}
{This specific choice leads to a \emph{complex multivariate Student-t} distribution for $\bm{x}_i$ after marginalizing over $\tau_i$ \cite{ollila2012complex,kotz2004multivariate}. While this defines a particular subclass of SIRVs, the derivations that follow provide a computationally efficient framework for robust covariance estimation and imputation within this class. The hyperparameters $\alpha_0$ and $\beta_0$ can be set to fixed values or estimated from the data.}

\subsection{Complete-Data Likelihood}
{When the inverse-gamma prior \eqref{eq:tau-prior} is adopted, the complete-data log-likelihood becomes, up to additive constants independent of $\bm{\Sigma}$ and $\tau_i$:}
\begin{equation}\label{eq:log-complete-ig}
\begin{split}
\log p(\bm{x}_i,\tau_i \condi \bm{\Sigma})
&= -l\log\tau_i - \log\Det{\bm{\Sigma}} - \frac{1}{\tau_i} \bm{x}_i^{\HH}\bm{\Sigma}^{-1}\bm{x}_i \\
&\quad - (\alpha_0+1)\log\tau_i - \frac{\beta_0}{\tau_i} + \mathrm{const}.
\end{split}
\end{equation}
{Grouping the terms by powers of $\tau_i$ yields the form used in the subsequent EM derivation.}

\subsection{EM Framework and Q-Function}

The Expectation-Maximization (EM) algorithm is an iterative procedure that
alternates between two steps: the Expectation (E) step, which computes the
expected complete-data log-likelihood given the current parameter estimate,
and the Maximization (M) step, which updates the parameters by maximizing
this expected log-likelihood.

At iteration $(t+1)$, the E-step computes the expectation of the
complete-data log-likelihood with respect to the posterior distribution of
the latent variables $\{\bm{x}_{i,m}, \tau_i\}$ conditioned on the observed
data $\bm{x}_{i,o}$ and the current estimate $\bm{\Sigma}^{(t)}$.
This expectation defines the $Q$-function:
\begin{equation}\label{eq:Q-def}
  Q(\bm{\Sigma} ; \bm{\Sigma}^{(t)})
  = \sum_{i=1}^n
    \EE\!\left[
      \log p(\bm{x}_i,\tau_i \condi \bm{\Sigma})
    \condi \bm{x}_{i,o}, \bm{\Sigma}^{(t)}
    \right],
\end{equation}
where $\EE[\,\cdot \condi \bm{x}_{i,o}, \bm{\Sigma}^{(t)}]$ denotes
expectation under the posterior
$p(\bm{x}_{i,m}, \tau_i \condi \bm{x}_{i,o}, \bm{\Sigma}^{(t)})$.

Focusing on the dependence on $\bm{\Sigma}$ and using
\eqref{eq:log-complete-ig} together with linearity of expectation, we obtain
\begin{equation}\label{eq:Q-decomp}
  Q(\bm{\Sigma} ; \bm{\Sigma}^{(t)})
  = -n\log\Det\bm{\Sigma}
    - \tr\!\Bigl(\bm{\Sigma}^{-1}\sum_{i=1}^n \bm{S}_i^{(t)}\Bigr)
    + K^{(t)},
\end{equation}
where the \emph{sufficient matrix} $\bm{S}_i^{(t)}$ is defined as
\begin{equation}\label{eq:Si-def}
  \bm{S}_i^{(t)}
  := \EE\!\left[
       \frac{1}{\tau_i}\,\bm{x}_i \bm{x}_i^{\HH}
     \condi \bm{x}_{i,o}, \bm{\Sigma}^{(t)}
     \right].
\end{equation}
All dependence of $Q$ on $\bm{\Sigma}$ is captured by the two explicit terms;
the scalar $K^{(t)}$ collects constants that are independent of $\bm{\Sigma}$.
The superscript $(t)$ in $\bm{S}_i^{(t)}$ and $K^{(t)}$ indicates that these
quantities are evaluated at the current estimate $\bm{\Sigma}^{(t)}$.

\subsection{E-step: Computation of Sufficient Statistics}

To compute $\bm{S}_i^{(t)}$ in closed form, we apply the tower rule
(law of iterated expectation) in two possible orders:
\begin{align}
  \bm{S}_i^{(t)}
  &= \EE\!\left[
       \frac{1}{\tau_i}\,\bm{x}_i \bm{x}_i^\HH
     \condi \bm{x}_{i,o}, \bm{\Sigma}^{(t)}
     \right]
  \label{eq:Si-def-explicit} \\
  &= \EE\!\Bigl[
       \EE\!\bigl[
         \tfrac{1}{\tau_i}\,\bm{x}_i \bm{x}_i^\HH
       \condi \bm{x}_{i,o}, \tau_i, \bm{\Sigma}^{(t)}
       \bigr]
     \condi \bm{x}_{i,o}, \bm{\Sigma}^{(t)}
     \Bigr]
  \label{eq:tower-1} \\
  &= \EE\!\Bigl[
       \EE\!\bigl[
         \tfrac{1}{\tau_i}\,\bm{x}_i \bm{x}_i^\HH
       \condi \bm{x}_{i,o}, \bm{x}_{i,m}, \bm{\Sigma}^{(t)}
       \bigr]
     \condi \bm{x}_{i,o}, \bm{\Sigma}^{(t)}
     \Bigr].
  \label{eq:tower-2}
\end{align}
Both orderings are algebraically equivalent, but the ordering in
\eqref{eq:tower-1}—conditioning first on $\tau_i$ and then averaging over
$\tau_i$—turns out to be analytically convenient for the chosen SIRV model
with conjugate inverse-gamma prior.
The reason is twofold.
On one hand, given $\tau_i$ and $\bm{\Sigma}^{(t)}$, the missing block
$\bm{x}_{i,m}$ is conditionally Gaussian:
\begin{equation}\label{eq:cond-xm}
  \bm{x}_{i,m}\mid \bm{x}_{i,o},\tau_i,\bm{\Sigma}^{(t)}
  \;\sim\;
  \mathcal{CN}\!\bigl(\bm{\mu}_i^{(t)},\,\tau_i\bm{C}_i^{(t)}\bigr),
\end{equation}
with conditional mean and conditional covariance given, respectively, by
\begin{equation}
  \bm{\mu}_i^{(t)}
  = \bm{\Sigma}_{mo}^{(t)}\,\bigl(\bm{\Sigma}_{oo}^{(t)}\bigr)^{-1}\bm{x}_{i,o},
  \qquad
  \bm{C}_i^{(t)}
  = \bm{\Sigma}_{mm}^{(t)}
    - \bm{\Sigma}_{mo}^{(t)}\,\bigl(\bm{\Sigma}_{oo}^{(t)}\bigr)^{-1}
      \bm{\Sigma}_{om}^{(t)}.
\end{equation}
This Gaussian structure allows the inner expectation in \eqref{eq:tower-1}
to be evaluated in closed form via standard second-moment identities.
On the other hand, under the inverse-gamma prior the posterior
$\tau_i \condi \bm{x}_{i,o}, \bm{\Sigma}^{(t)}$ remains inverse-gamma
(see Section~\ref{sec:outer-exp}), so the outer expectation reduces to a
simple scalar multiplication.
By contrast, the ordering in \eqref{eq:tower-2} would require computing
$\EE_{\bm{x}_{i,m}}\!\bigl[(\beta_0 +
\bm{x}_i^\HH(\bm{\Sigma}^{(t)})^{-1}\bm{x}_i)^{-1}\bigr]$,
which is not available in closed form.

\subsection{Inner Expectation Conditioned on $\tau_i$}
\label{sec:inner-exp}

For a fixed $\tau_i > 0$, we define
\begin{equation}
  A_i(\tau_i)
  := \EE\!\left[
       \frac{1}{\tau_i}\,\bm{x}_i \bm{x}_i^\HH
     \condi \bm{x}_{i,o}, \tau_i, \bm{\Sigma}^{(t)}
     \right].
\end{equation}
Using \eqref{eq:cond-xm} and the identity
$\EE[\bm{z}\bm{z}^\HH] = \mathrm{Cov}(\bm{z}) + \EE[\bm{z}]\EE[\bm{z}]^\HH$
for a random vector $\bm{z}$, the matrix $A_i(\tau_i)$ takes the block form
\begin{equation}\label{eq:A-blocks}
  A_i(\tau_i)
  = \begin{pmatrix}
      \tfrac{1}{\tau_i}\,\bm{x}_{i,o}\bm{x}_{i,o}^\HH
      & \tfrac{1}{\tau_i}\,\bm{x}_{i,o}(\bm{\mu}_i^{(t)})^\HH \\[4pt]
      \tfrac{1}{\tau_i}\,\bm{\mu}_i^{(t)}\bm{x}_{i,o}^\HH
      & \bm{C}_i^{(t)}
        + \tfrac{1}{\tau_i}\,\bm{\mu}_i^{(t)}(\bm{\mu}_i^{(t)})^\HH
    \end{pmatrix}.
\end{equation}
Here $A_i(\tau_i)$ is partitioned conformably with the decomposition of
$\bm{x}_i = (\bm{x}_{i,o}^\top, \bm{x}_{i,m}^\top)^\top$ into its observed
part $\bm{x}_{i,o}$ (indexed by $\mathcal{O}_i$) and its missing part
$\bm{x}_{i,m}$ (indexed by $\mathcal{M}_i$).
The top-left entry, corresponding to the observed--observed subspace, follows
immediately since $\bm{x}_{i,o}$ is fixed and contributes no randomness.
The top-right entry, corresponding to the observed--missing subspace, equals
$\bm{x}_{i,o}\,\EE[\bm{x}_{i,m}^\HH \condi
\bm{x}_{i,o},\tau_i,\bm{\Sigma}^{(t)}]/\tau_i
= \bm{x}_{i,o}(\bm{\mu}_i^{(t)})^\HH/\tau_i$,
because $\bm{x}_{i,o}$ is fixed and the conditional mean of $\bm{x}_{i,m}$
is $\bm{\mu}_i^{(t)}$ by \eqref{eq:cond-xm}.
Finally, for the bottom-right entry corresponding to the missing--missing
subspace, the second-moment identity gives
$\EE[\bm{x}_{i,m}\bm{x}_{i,m}^\HH \condi
\bm{x}_{i,o},\tau_i,\bm{\Sigma}^{(t)}]
= \tau_i\bm{C}_i^{(t)} + \bm{\mu}_i^{(t)}(\bm{\mu}_i^{(t)})^\HH$,
and dividing by $\tau_i$ yields the bottom-right expression in
\eqref{eq:A-blocks}.

\subsection{Outer Expectation: Posterior of $\tau_i$ and Final Form}
\label{sec:outer-exp}

Under the inverse-gamma prior \eqref{eq:tau-prior}, a standard Bayesian
conjugacy argument shows that the posterior of $\tau_i$ given the observed
block $\bm{x}_{i,o}$ and the current estimate $\bm{\Sigma}^{(t)}$ is again
inverse-gamma:
\begin{equation}\label{eq:tau-posterior}
  \tau_i \condi \bm{x}_{i,o}, \bm{\Sigma}^{(t)}
  \;\sim\;
  \IG\!\bigl(\alpha_0 + d_i,\;\beta_0 + q_i^{(t)}\bigr),
\end{equation}
where $d_i = |\mathcal{O}_i|$ is the number of observed components of
$\bm{x}_i$ and
\begin{equation}\label{eq:qi-def}
  q_i^{(t)}
  := \bm{x}_{i,o}^\HH\,\bigl(\bm{\Sigma}_{oo}^{(t)}\bigr)^{-1}\bm{x}_{i,o}
\end{equation}
is the Mahalanobis distance of the observed block.
From the mean of the inverse-gamma distribution, the scalar weight needed in
the E-step is therefore
\begin{equation}\label{eq:a-def}
  a_i^{(t)}
  := \EE\!\left[
       \frac{1}{\tau_i}
     \condi \bm{x}_{i,o}, \bm{\Sigma}^{(t)}
     \right]
  = \frac{\alpha_0 + d_i}{\beta_0 + q_i^{(t)}},
\end{equation}
which is fully computable from the observed data and the current estimate.
Combining \eqref{eq:A-blocks} and \eqref{eq:a-def}, the sufficient matrix
takes the explicit closed form
\begin{equation}\label{eq:Si-final}
  \bm{S}_i^{(t)}
  = \begin{pmatrix}
      a_i^{(t)}\,\bm{x}_{i,o}\bm{x}_{i,o}^\HH
      & a_i^{(t)}\,\bm{x}_{i,o}(\bm{\mu}_i^{(t)})^\HH \\[4pt]
      a_i^{(t)}\,\bm{\mu}_i^{(t)}\bm{x}_{i,o}^\HH
      & \bm{C}_i^{(t)}
        + a_i^{(t)}\,\bm{\mu}_i^{(t)}(\bm{\mu}_i^{(t)})^\HH
    \end{pmatrix},
\end{equation}
which is the central quantity passed to the M-step.

\subsection{M-step: Covariance Update}

In the M-step, the covariance matrix is updated by maximizing the
$Q$-function with respect to $\bm{\Sigma}$.
Substituting \eqref{eq:Si-final} into \eqref{eq:Q-decomp} and
differentiating with respect to $\bm{\Sigma}$ via standard matrix-calculus
identities, the stationarity condition reads
\begin{equation}
  -n\bm{\Sigma}^{-1}
  + \bm{\Sigma}^{-1}\!\Bigl(\sum_{i=1}^n \bm{S}_i^{(t)}\Bigr)\bm{\Sigma}^{-1}
  = \bm{0}.
\end{equation}
Multiplying both sides on the left and right by $\bm{\Sigma}$ immediately
yields the explicit update rule
\begin{equation}\label{eq:Sigma-update}
  \bm{\Sigma}^{(t+1)}
  = \frac{1}{n}\sum_{i=1}^n \bm{S}_i^{(t)}.
\end{equation}
The M-step therefore reduces to a simple empirical average of the sufficient
matrices $\bm{S}_i^{(t)}$ computed during the E-step, with no iterative
inner loop required.


\section{Experimental Validation}
\label{sec:simulation}

We evaluate the performance of several covariance estimation and imputation methods on synthetic complex-valued data with controlled missingness under MCAR mechanisms. All Monte Carlo experiments are ran with a fixed random seed for reproducibility. Unless otherwise stated, the main experimental settings are: number of Monte Carlo runs $MC=100$, dimension $l=8$, number of samples $n=1000$.

\subsection{Synthetic datasets}
\label{ssec:synthetic-datasets}

Synthetic observations $\bm{X}\in\mathbb{C}^{n\times l}$ are generated from a complex Gaussian-like model built from a true covariance matrix $\bm{\Sigma}_{\mathrm{true}}$, which is a Toeplitz matrix with correlation parameter $\rho=0.7$. For each Monte Carlo repetition we draw $n$ independent samples using the Cholesky factor of $\bm{\Sigma}_{\mathrm{true}}$. An independent scale factor vector $\tau$ is set to ones by default (the Student-$t$/SIRV heavy-tailed extension can be activated if needed).

To assess robustness, we inject two types of data degradation. First, a missing completely at random (MCAR) mechanism masks a fraction \texttt{missRate = 0.30} of the entries of the first data block $X_1$; the masked entries are set to \texttt{NaN} and are later used to evaluate imputation accuracy, while the ground truth values are preserved from the generated complete dataset $X$. Second, row-wise outliers are introduced by corrupting a fraction \texttt{outlier\_frac = 0.1} of rows, either additively or by replacement. In the additive mode, an outlier matrix $X_2$ (with amplitude scaled by \texttt{amp\_factor}) is added to the selected rows; otherwise, the rows are replaced by $X_2$. Outliers are injected only to fully observed rows.

We compare five different methods for covariance estimation under missing data. The first, EM-SIRV, is the robust Expectation-Maximization algorithm specifically adapted to SIRV models as derived in this paper. The second, EM-EOF \cite{hippert2020eof}, combines the EM framework with an EOF/PCA-based low-rank approximation, allowing the optimal rank $k$ to be selected. The third, MeanImpute, performs a simple column-wise mean imputation before computing the empirical covariance. The fourth, PCA imputation \cite{josse2012handling}, reconstructs the data through a low-rank PCA model with a chosen number of components. Finally, EM-Gauss corresponds to the classical EM algorithm designed for Gaussian data with missing values. For all methods, the estimated covariance matrices are symmetrized and projected onto the nearest Hermitian positive semi-definite matrix before computing the performance metrics.

To evaluate the quality of the imputation and covariance estimation, a set of complementary metrics is employed. Denoting by $\hat{\bm{x}}_{i,j}$ the imputed value of the $j$-th component of sample $i$ (after unscaling, if applicable) and by $\bm{x}_{i,j}$ the corresponding true value, the normalized mean squared error (NMSE) is defined as
\begin{equation}
\mathrm{NMSE} = \frac{\sum_{(i,j)\in\mathcal{M}} |\hat{\bm{x}}_{i,j} - \bm{x}_{i,j}|^2}{\sum_{(i,j)\in\mathcal{M}} |\bm{x}_{i,j}|^2},
\end{equation}
where $\mathcal{M}$ is the set of masked entries. The complex root-mean-square error (RMSE-cpx) is given by $\sqrt{\frac{1}{|\mathcal{M}|}\sum_{\mathcal{M}} |\hat{\bm{x}}_{i,j} - \bm{x}_{i,j}|^2}$. For applications involving interferometric phase and amplitude, the phase RMSE (in radians) and amplitude RMSE are evaluated separately by decomposing the complex residuals. The accuracy of the covariance estimate $\hat{\bm{\Sigma}}$ is measured by the Frobenius distance $\|\hat{\bm{\Sigma}} - \bm{\Sigma}_{\mathrm{true}}\|_F$. All metrics are averaged over Monte Carlo repetitions; boxplots summarize their distributions across runs.

\subsubsection*{Results on synthetic data}
\begin{figure*}[!t]
  \centering

  \begin{subfigure}[b]{0.45\textwidth}
    \centering
    \includegraphics[width=\linewidth]{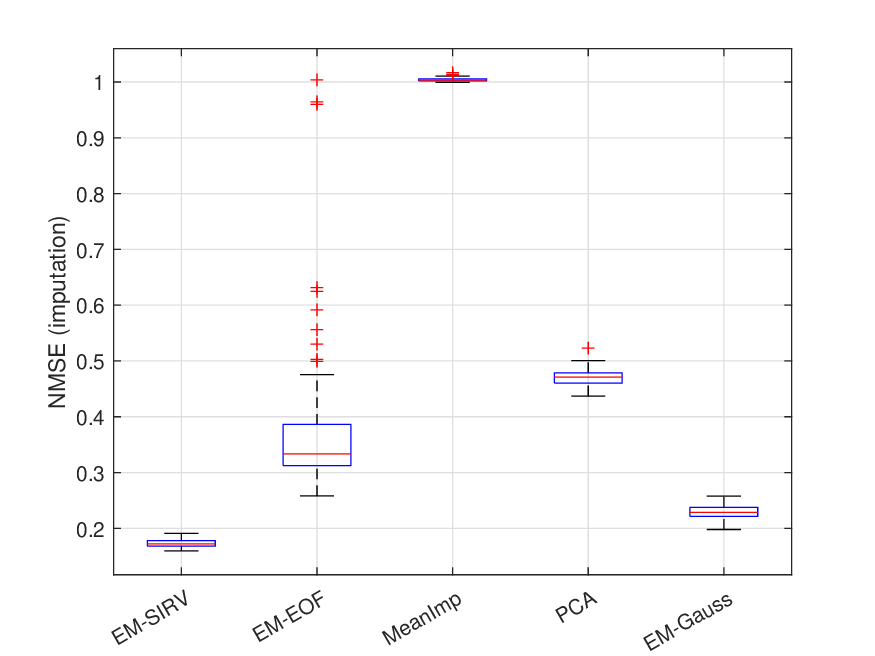}
    \caption{Boxplots of NRMSE for imputation across Monte Carlo repetitions.}
    \label{fig:imputation}
  \end{subfigure}\hfill
  \begin{subfigure}[b]{0.45\textwidth}
    \centering
    \includegraphics[width=\linewidth]{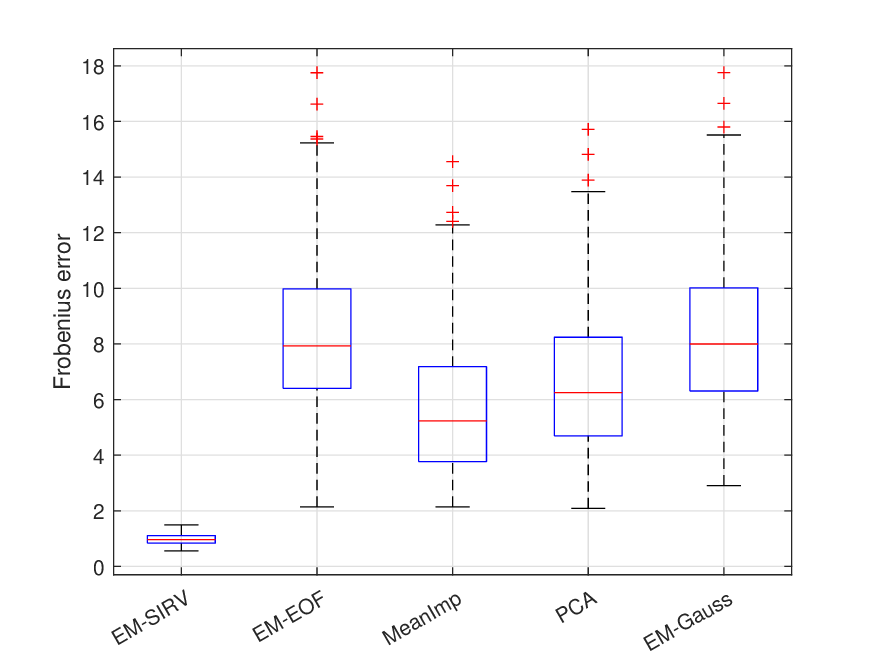}
    \caption{Boxplots of the Frobenius distance between estimated and true covariance matrices.}
    \label{fig:frob}
  \end{subfigure}
  
  \vspace{0.3cm}
  \begin{subfigure}[b]{0.45\textwidth}
    \centering
    \includegraphics[width=\linewidth]{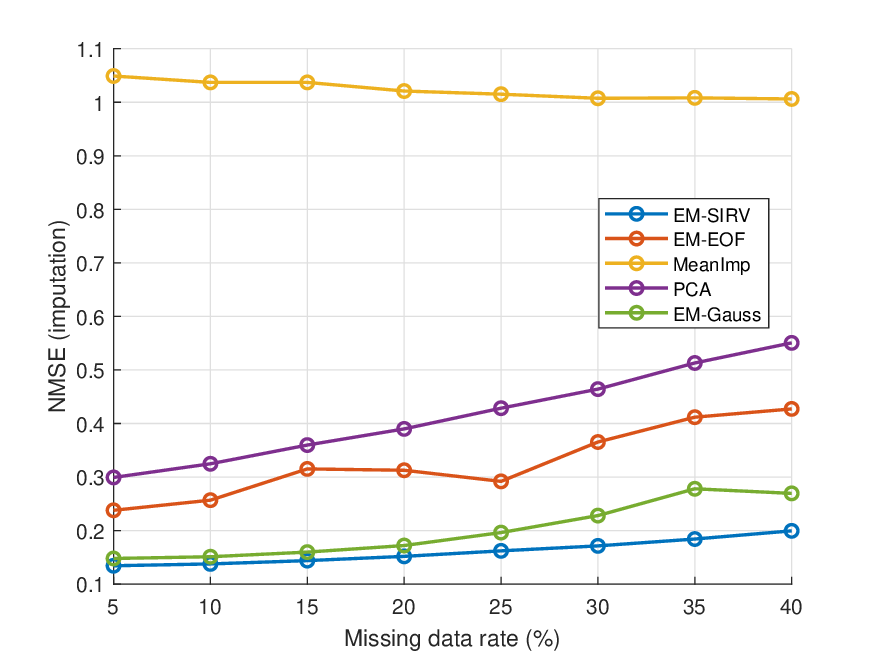}
    \caption{Mean NMSE (imputation) vs.\ missing data rate (fixed outlier rate = 20\%).}
    \label{fig:imputation_miss}
  \end{subfigure}\hfill
  \begin{subfigure}[b]{0.45\textwidth}
    \centering
    \includegraphics[width=\linewidth]{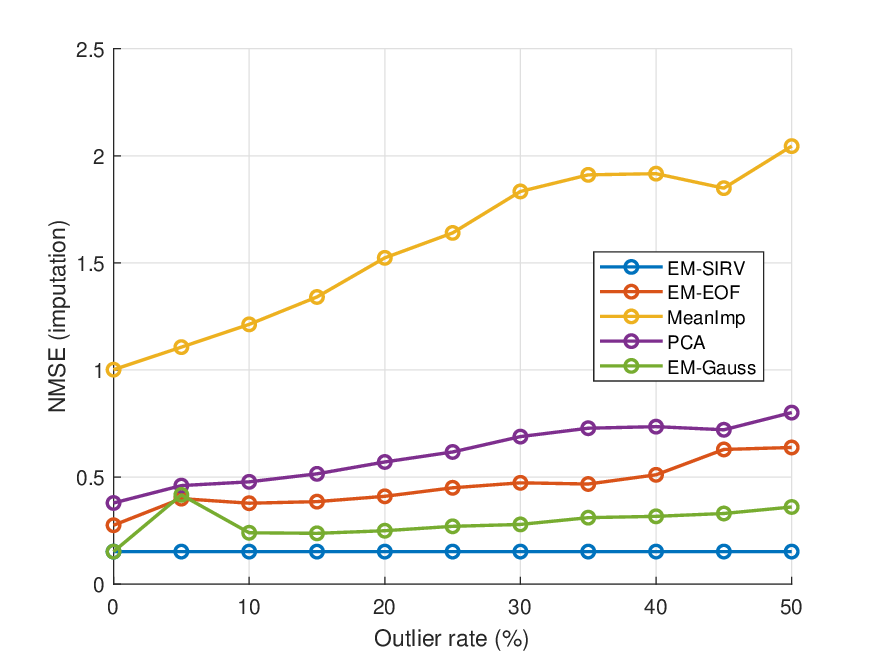}
    \caption{Mean NMSE (imputation) vs.\ outlier rate (fixed missing rate = 20\%).}
    \label{fig:imputation_out}
  \end{subfigure}
  
  \caption{Performance comparison on synthetic data with MCAR missing entries and row-wise outliers. (a, b) Results for 30\% missing rate and 10\% outliers (Monte Carlo repetitions). (c, d) Mean NMSE under varying missing rate (5\% to 40\%) with 20\% outliers, and varying outlier rate (0\% to 50\%) with 20\% missing entries.}
  \label{fig:combined}
\end{figure*}

Figures\ref{fig:imputation} and \ref{fig:frob} show the performance comparison on synthetic data. The proposed EM-SIRV method consistently outperforms the other approaches in both imputation accuracy and covariance estimation, particularly in the presence of outliers and non-Gaussian characteristics. The robustness of the SIRV model to heavy-tailed distributions is evident in these results.
Figures ~\ref{fig:imputation_miss} and ~\ref{fig:imputation_out} compares the imputation NMSE of all methods under two scenarios: varying missing data rate with fixed 20\% outliers, and varying outlier rate with fixed 20\% missing entries. In both cases, EM‑SIRV achieves the lowest NMSE across the whole range. EM‑Gauss ranks second, followed by EM‑EOF and PCA, while mean imputation shows the highest errors. The performance gap widens with increasing corruption, demonstrating the robustness of the proposed SIRV‑based EM algorithm.

\subsection{Real Dataset}
\label{sec:real_data}

We evaluate the proposed method on a real Sentinel-1 InSAR dataset over
Mexico City, a region affected by significant land subsidence due to
intensive groundwater extraction. The data consist of a complex-valued
cube of dimensions $1000 \times 2200 \times 10$, where the three axes
correspond to rows, columns, and interferometric acquisitions,
respectively. The 10 interferograms were acquired every 12 days over
the period from August 14, 2019 to November 18, 2019.

For the experiments, the interferometric cube is reshaped into a matrix
$\bm{X}\in\mathbb{C}^{n\times l}$, with
$n = 1{,}000 \times 2{,}200 = 2{,}200{,}000$ pixels and $l = 10$
interferograms. Each pixel time series is normalized by its mean
amplitude across acquisitions to reduce scale variability across
spatial locations. The complete cube serves both as input and as
ground truth to assess reconstruction quality and covariance estimation
accuracy.

To evaluate the robustness of the proposed method under realistic
degradation scenarios, three missing-data mechanisms are considered.
The first two are MAR mechanisms applied with a fixed overall missing
rate of $25\%$; the third is an MNAR mechanism based on coherence
thresholding. In addition, $10\%$ of all pixels are simultaneously
corrupted by amplitude outliers in all three scenarios.

\subsubsection*{Missing Data Mechanisms}

\textbf{Spatial patches.}
Spatially coherent missing regions are generated to simulate localized
decorrelation zones such as vegetated areas or high-phase-noise urban
patches. At each acquisition $t$, a white-noise field is convolved with
a Gaussian kernel of standard deviation
$\sigma_{\mathrm{sp}} = 0.02\min(H,W)$ pixels and thresholded at its
$(1-p)$-th percentile to produce the binary mask. A persistence
parameter $\gamma = 0.20$ carries a random fraction of the previous
mask forward, introducing mild temporal continuity between consecutive
acquisitions while keeping the dominant structure spatially driven.

\textbf{Spatio-temporal patches.}
Spatially and temporally correlated missing regions are produced to
mimic persistent atmospheric disturbances or slowly evolving
decorrelation fronts. A latent spatial field $Z_t$ follows a
first-order autoregressive process,
\begin{equation}\label{eq:ar1}
Z_t = \rho\, Z_{t-1} + \sqrt{1-\rho^2}\,\varepsilon_t,
\qquad \varepsilon_t \sim \mathcal{N}(\bm{0},\bm{I}),
\end{equation}
with temporal correlation $\rho = 0.85$. At each step, $Z_t$ is
smoothed by a Gaussian kernel of standard deviation
$\sigma_{\mathrm{sp}} = 0.03\min(H,W)$ pixels and thresholded at its
$(1-p)$-th percentile. The high value of $\rho$ produces slowly
evolving connected patches that are strongly correlated across
consecutive interferograms, representing the most challenging MAR
scenario for imputation.

\textbf{MNAR coherence-based self-masking.}
To assess the proposed method under a non-ignorable missingness
mechanism, we consider a scenario in which pixels whose interferometric
coherence falls below a threshold $\gamma = 0.40$ are declared missing.
This self-masking mechanism is physically motivated---low-coherence
pixels are genuinely unreliable in InSAR---and statistically
challenging, since the missingness probability is correlated with the
signal quality itself. For each slave acquisition $t \in \{2,\ldots,T\}$,
the interferometric coherence with respect to the master (acquisition $t=1$)
is estimated over a $5\times5$ sliding spatial window:
\begin{equation}\label{eq:coh}
  \hat{\gamma}_{t} =
  \frac{\bigl|\sum_{\mathcal{W}} z_{t}\,z_{1}^{*}\bigr|}
       {\sqrt{\sum_{\mathcal{W}}|z_{t}|^2 \;\sum_{\mathcal{W}}|z_{1}|^2}},
\end{equation}
where $\mathcal{W}$ denotes the local spatial window, $z_t$ the complex
SLC value at acquisition $t$, and $z_1$ the master SLC. A pixel is then
masked at all dates if its minimum master--slave coherence falls below
the threshold:
\begin{equation}\label{eq:mask}
  \min_{t \in \{2,\ldots,T\}} \hat{\gamma}_{t} < \gamma,
\end{equation}
yielding a single spatial mask applied uniformly across acquisitions,
with approximately $13.8\%$ of missing entries concentrated in
low-coherence zones (vegetated areas, layover).
\subsubsection*{Outlier Model}

A fraction of $10\%$ of all $n$ pixels are corrupted by amplitude
spikes. Outlier pixels are drawn exclusively from the set of
fully-observed pixels (those with no missing entry across all $l$
acquisitions), ensuring that the outlier and missing-data masks are
disjoint by construction. Their entire temporal profile is multiplied
by a random factor $f_i \sim \mathcal{U}[60,\,100]$:
\begin{equation}\label{eq:outlier-model}
\tilde{\bm{x}}_i = f_i\, \bm{x}_i,
\qquad f_i \sim \mathcal{U}[60,\,100],
\end{equation}
where $\bm{x}_i \in \mathbb{C}^l$ is the normalized pixel time series.
This model mimics strong double-bounce or man-made reflector returns
common in urban InSAR scenes~\cite{chen2018robust,bamler1998synthetic,
zebker1992decorrelation} and constitutes a severe test for robust
estimators, since the corrupted pixels retain their original phase
structure while dominating the empirical covariance through their
inflated amplitude. This outlier injection procedure is applied
identically across all three scenarios. For visualization purposes,
outlier pixels are replaced by their ground-truth values in all
displayed phase maps, so that visual comparisons focus exclusively on
imputation quality at the masked locations.

\subsubsection*{Experimental Setup}

All three mechanisms are evaluated under identical conditions (outlier
rate $10\%$, fixed random seed) to ensure a fair comparison.
Performance on the masked entries is measured by four complementary
metrics: the normalized mean squared error (NMSE) on the complex
residuals, the phase RMSE (in radians), the complex RMSE, and the
Signal-to-Distortion Ratio (SDR in dB).
The preprocessing of the Sentinel-1 data was performed using the
Sentinel Application Platform (SNAP) developed by the European Space
Agency~\cite{esa_snap}. The processing chain is available
online\footnote{\href{https://github.com/DanaElhajjar/InSAR_ChainProcessing}%
{https://github.com/DanaElhajjar/InSAR\_ChainProcessing}},
and the dataset is publicly
available\footnote{\href{https://doi.org/10.5281/zenodo.11283419}%
{10.5281/zenodo.11283419}}.

\subsubsection*{Results on Real Data}

\begin{figure*}[!t]
    \centering
    \begin{subfigure}[t]{0.95\textwidth}
        \centering
        \includegraphics[width=\linewidth]{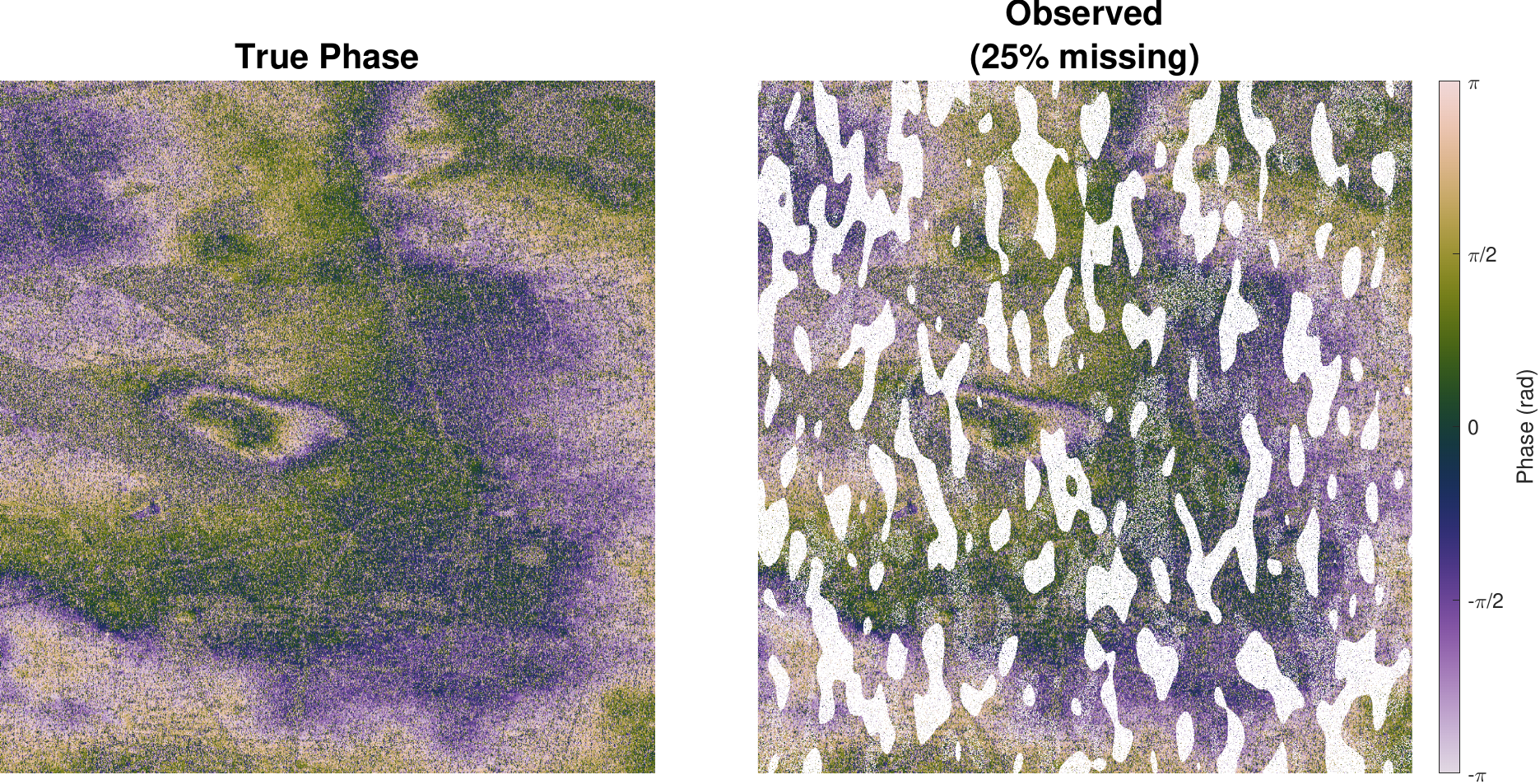}
        \caption{True and observed phase for acquisition $T=7$.}
        \label{fig:t7_phase_true_obs_SP}
    \end{subfigure}
    \vspace{0.4cm}
    \begin{subfigure}[t]{0.95\textwidth}
        \centering
        \includegraphics[width=\linewidth]{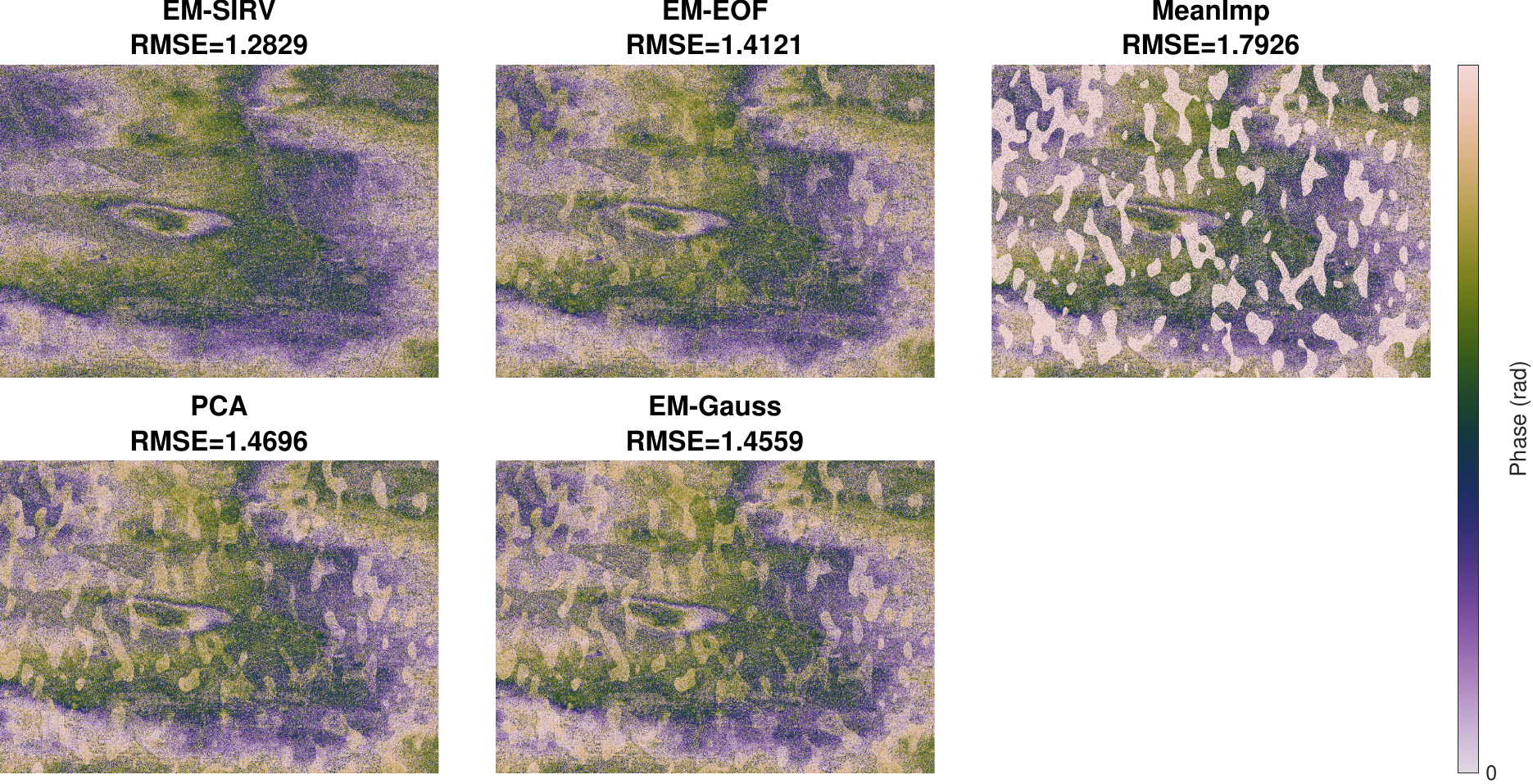}
        \caption{Phase estimates of the methods for acquisition $T=7$.}
        \label{fig:t7_phase_methods_SP}
    \end{subfigure}
    \caption{Results for acquisition $T=7$: phase comparisons with 25\%
    spatial patches gaps and 10\% outliers.}
    \label{fig:results_t7_SP}
\end{figure*}

\begin{figure*}[!t]
    \centering
    \begin{subfigure}[t]{0.95\textwidth}
        \centering
        \includegraphics[width=\linewidth]{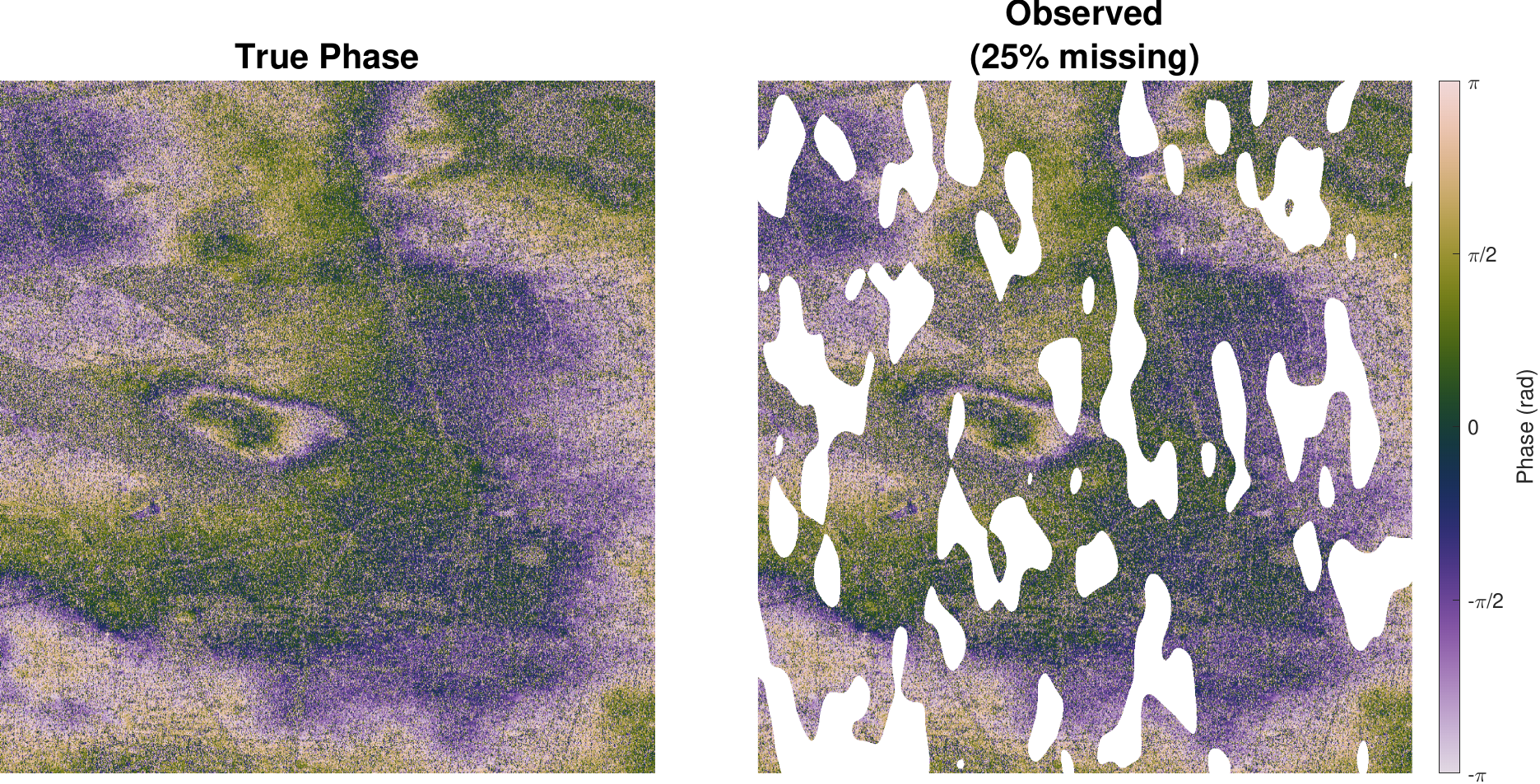}
        \caption{True and observed phase for acquisition $T=7$.}
        \label{fig:t7_phase_true_obs_SPT}
    \end{subfigure}
    \vspace{0.4cm}
    \begin{subfigure}[t]{0.95\textwidth}
        \centering
        \includegraphics[width=\linewidth]{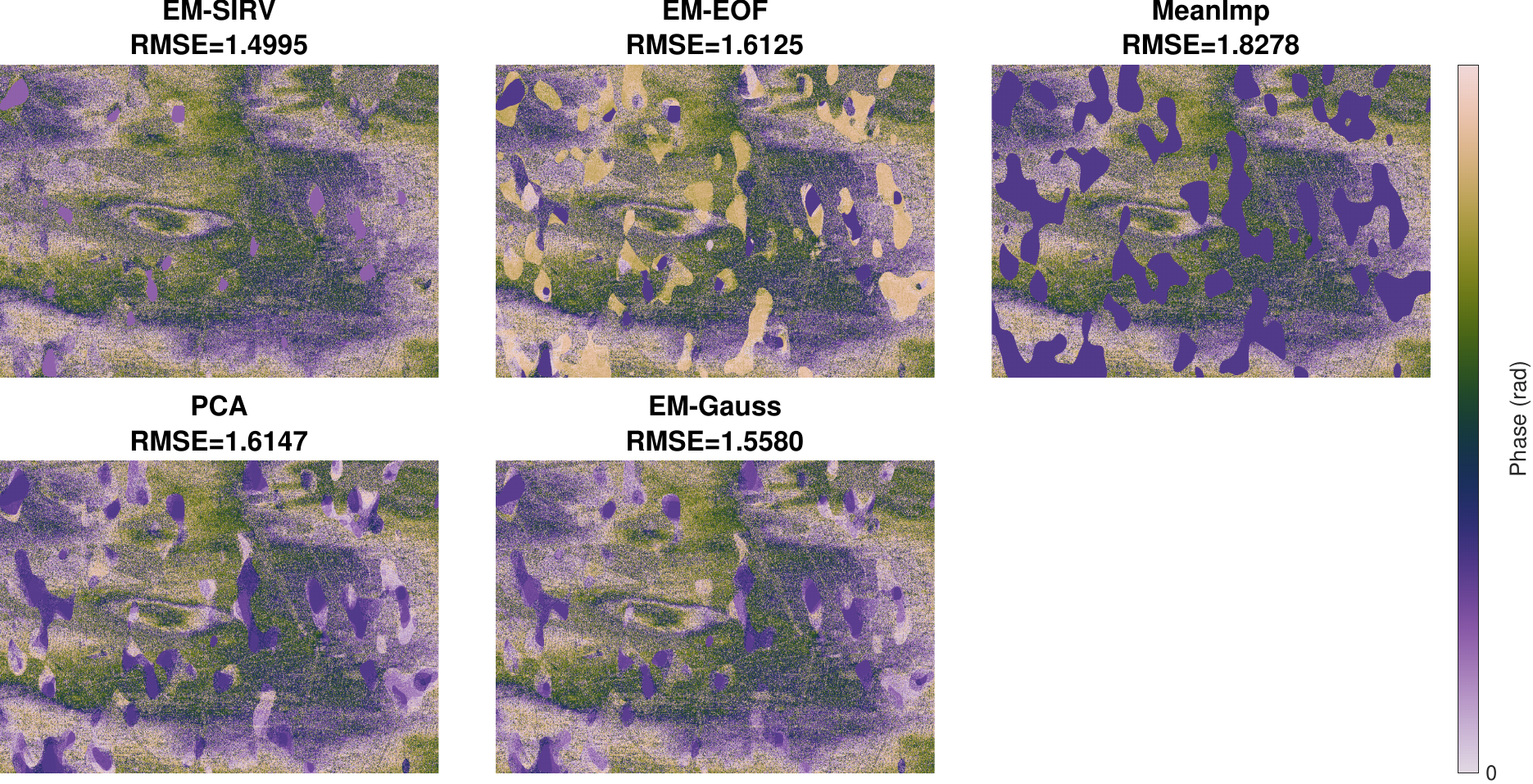}
        \caption{Phase estimates of the methods for acquisition $T=7$.}
        \label{fig:t7_phase_methods_SPT}
    \end{subfigure}
    \caption{Results for acquisition $T=7$: phase comparisons with 25\%
    spatio-temporal patches gaps and 10\% outliers.}
    \label{fig:results_t7_SPT}
\end{figure*}

Figures~\ref{fig:results_t7_SP} and~\ref{fig:results_t7_SPT} present
visual results for acquisition $T=7$ under the spatial patches and
spatio-temporal patches scenarios, respectively, with a missing rate of
$25\%$ and $10\%$ amplitude outliers. Quantitative results, aggregated
over all masked entries across the ten acquisitions, are reported in
Tables~\ref{tab:mexico_spatial} and~\ref{tab:mexico_spatiotemporal}.

Under the spatial patches scenario, EM-SIRV achieves the best
performance across all metrics, with a complex RMSE of $1.009$, a
phase RMSE of $1.339$ rad, an NMSE of $0.845$, and an SDR of $0.73$ dB.
All competing methods yield higher NMSE and negative SDR values,
indicating that their reconstructions are of lower quality than the
trivial mean baseline. Visually, the phase maps reconstructed by
EM-SIRV for $T=7$ (Figure~\ref{fig:results_t7_SP}) faithfully restore
the spatial structure of the interferograms, including the
characteristic subsidence fringes over Mexico City, while competing
methods introduce artefacts or over-smooth spatial details in regions
with dense missing data.

Under the more challenging spatio-temporal patches scenario, the
performance gap widens further. EM-SIRV retains competitive performance
with a complex RMSE of $1.048$, a phase RMSE of $1.479$ rad, an NMSE
of $0.909$, and an SDR of $0.41$ dB, whereas the other methods exhibit
SDR values ranging from $-3.61$ dB (EM-Gauss) to $-7.74$ dB (EM-EOF).
The particularly severe degradation of EM-EOF under strongly correlated
gaps highlights its sensitivity to the temporal structure of the
missing-data patterns. Qualitatively, the reconstructions produced by
EM-SIRV for $T=7$ (Figure~\ref{fig:results_t7_SPT}) preserve the
spatial coherence of the interferometric fringes despite the extent and
temporal persistence of the masked regions, whereas competing methods
produce incoherent or heavily distorted maps in the affected areas.

\begin{figure*}[!t]
    \centering
    \begin{subfigure}[t]{0.95\textwidth}
        \centering
        \includegraphics[width=\linewidth]{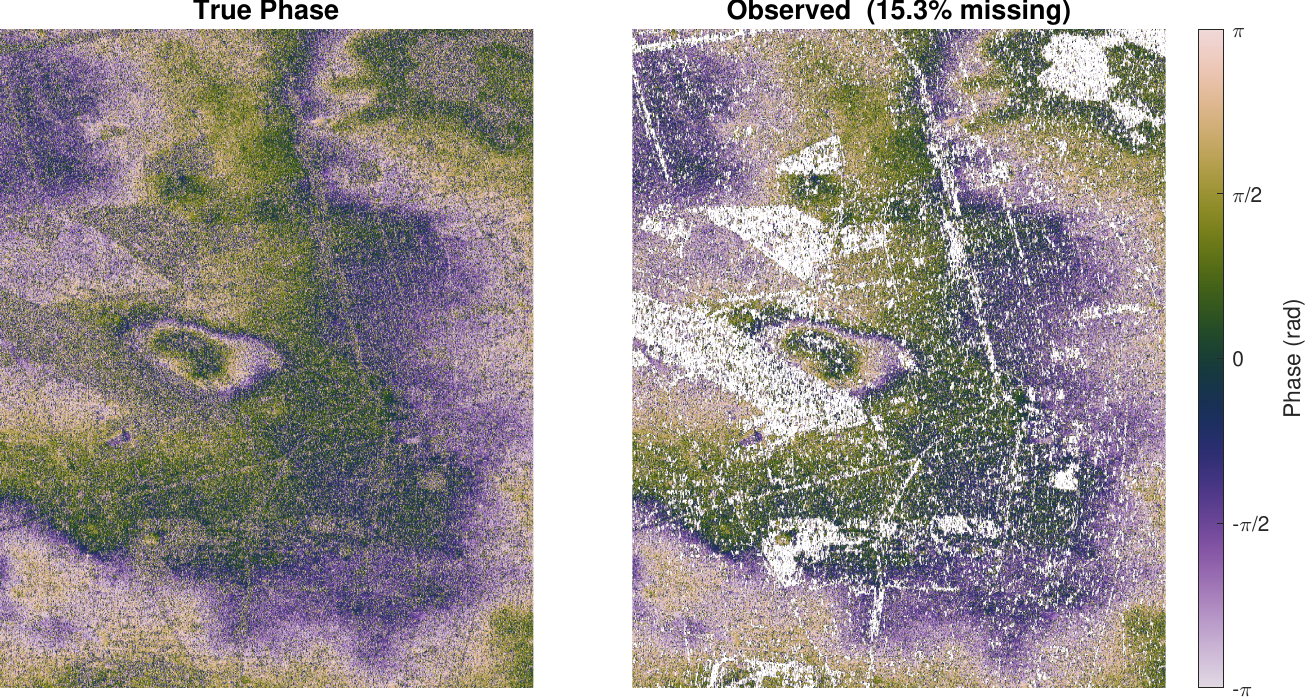}
        \caption{True and observed phase for acquisition $T=7$.}
        \label{fig:t7_phase_true_obs_mnar}
    \end{subfigure}
    \vspace{0.4cm}
    \begin{subfigure}[t]{0.95\textwidth}
        \centering
        \includegraphics[width=\linewidth]{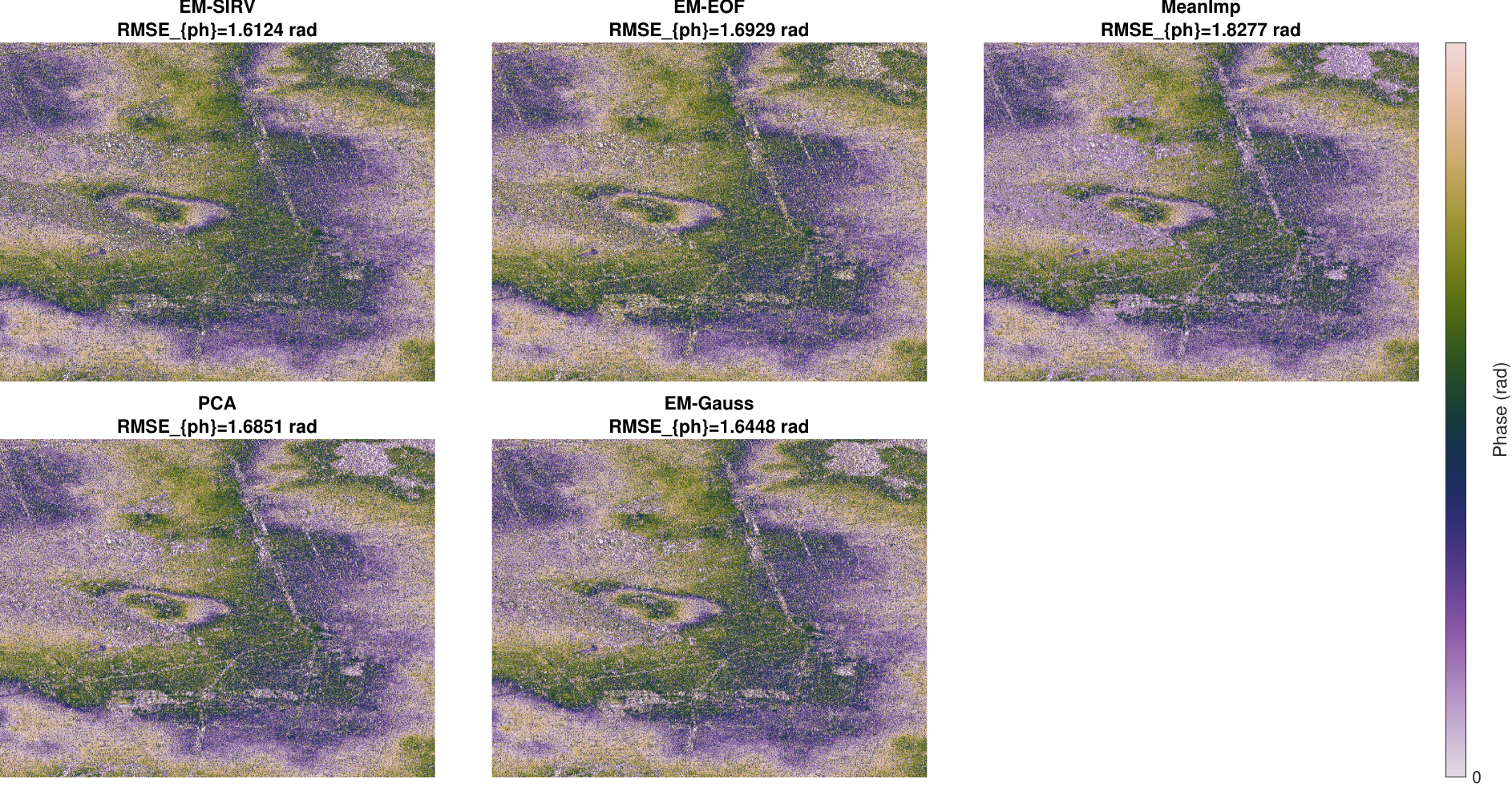}
        \caption{Phase estimates of the methods for acquisition $T=7$.}
        \label{fig:t7_phase_methods_mnar}
    \end{subfigure}
    \caption{Results for acquisition $T=7$: phase comparisons with a
    coherence threshold $\gamma=0.40$, resulting in $13.8\%$ missing data
    under MNAR self-masking and $10\%$ outliers.}
    \label{fig:results_t7_mnar}
\end{figure*}

Under the MNAR self-masking scenario, quantitative results are reported
in Table~\ref{tab:mexico_coherence_selfmasking}.
Figure~\ref{fig:results_t7_mnar}(a) displays, for acquisition $T=7$,
the true phase and the observed phase after coherence-based masking;
the masked regions clearly correspond to low-coherence zones (vegetated
areas and layover), confirming the self-masking nature of the
mechanism. Figure~\ref{fig:results_t7_mnar}(b) compares the phase
reconstructions produced by all five methods. EM-SIRV visually
preserves the characteristic subsidence fringes in the masked regions,
whereas competing methods introduce spatial incoherence or over-smooth
the fringe pattern, a degradation most pronounced for EM-EOF and
MeanImp. Despite the MNAR nature of the missingness---which formally
violates the MAR assumption underlying our derivation---EM-SIRV
consistently achieves the best performance across all metrics: complex
RMSE of $1.148$, phase RMSE of $1.630$ rad, NMSE of $1.026$, and SDR
of $-0.11$~dB. All competing methods yield substantially higher errors
(SDR between $-2.69$~dB for EM-Gauss and $-5.76$~dB for EM-EOF).
This behavior is consistent with the observation that coherence-based
self-masking tends to remove the least informative pixels, making the
remaining observed data relatively more representative of the
underlying signal distribution.

These results confirm the robustness of the proposed EM-SIRV framework
to both heavy-tailed non-Gaussian distributions and diverse
missing-data mechanisms, including MNAR regimes, making it well suited
for operational InSAR time series processing in urban environments.


\begin{table}[!t]
\centering
\setlength{\tabcolsep}{4pt}
\footnotesize
\caption{Quantitative results on the Sentinel-1 Mexico City dataset
under the \textbf{spatial patches} scenario
(missing rate 25\%, outlier rate 10\%).
Best results are highlighted in \textbf{bold}.}
\label{tab:mexico_spatial}
\renewcommand{\arraystretch}{1.2}
\begin{tabular}{lcccc}
\toprule
\textbf{Method} & \textbf{NMSE} & \textbf{RMSE-cpx}
                & \textbf{RMSE-ph (rad)} & \textbf{SDR (dB)} \\
\midrule
EM-SIRV  & \textbf{0.8447} & \textbf{1.0085} & \textbf{1.3393} & \textbf{0.73} \\
EM-EOF   & 1.5163          & 1.3512          & 1.4759          & -1.81         \\
MeanImp  & 2.1636          & 1.6140          & 1.6924          & -3.35         \\
PCA      & 1.4333          & 1.3137          & 1.4679          & -1.56         \\
EM-Gauss & 1.4200          & 1.3076          & 1.4499          & -1.52         \\
\bottomrule
\end{tabular}
\end{table}

\begin{table}[!t]
\centering
\setlength{\tabcolsep}{4pt}
\footnotesize
\caption{Quantitative results on the Sentinel-1 Mexico City dataset
under the \textbf{spatio-temporal patches} scenario
(missing rate 25\%, outlier rate 10\%).
Best results are highlighted in \textbf{bold}.}
\label{tab:mexico_spatiotemporal}
\renewcommand{\arraystretch}{1.2}
\begin{tabular}{lcccc}
\toprule
\textbf{Method} & \textbf{NMSE} & \textbf{RMSE-cpx}
                & \textbf{RMSE-ph (rad)} & \textbf{SDR (dB)} \\
\midrule
EM-SIRV  & \textbf{0.9090} & \textbf{1.0481} & \textbf{1.4787} & \textbf{0.41} \\
EM-EOF   & 5.9461          & 2.6806          & 1.6100          & -7.74         \\
MeanImp  & 3.2854          & 1.9926          & 1.7021          & -5.17         \\
PCA      & 2.3855          & 1.6979          & 1.5928          & -3.78         \\
EM-Gauss & 2.2976          & 1.6663          & 1.5694          & -3.61         \\
\bottomrule
\end{tabular}
\end{table}

\begin{table}[!t]
\centering
\setlength{\tabcolsep}{4pt}
\footnotesize
\caption{Quantitative results on the Sentinel-1 Mexico City dataset
under the \textbf{MNAR self-masking} scenario
(coherence threshold $\gamma=0.40$, resulting in $13.8\%$ missing data
and $10\%$ outliers).
Best results are highlighted in \textbf{bold}.}
\label{tab:mexico_coherence_selfmasking}
\renewcommand{\arraystretch}{1.2}
\begin{tabular}{lcccc}
\toprule
\textbf{Method} & \textbf{NMSE} & \textbf{RMSE-cpx}
                & \textbf{RMSE-ph (rad)} & \textbf{SDR (dB)} \\
\midrule
EM-SIRV  & \textbf{1.0256} & \textbf{1.1483} & \textbf{1.6302} & \textbf{-0.11} \\
EM-EOF   & 3.7648          & 2.2002          & 1.7055          & -5.76          \\
MeanImp  & 2.3628          & 1.7430          & 1.7612          & -3.73          \\
PCA      & 1.9510          & 1.5838          & 1.6804          & -2.90          \\
EM-Gauss & 1.8570          & 1.5452          & 1.6620          & -2.69          \\
\bottomrule
\end{tabular}
\end{table}

\section*{Conclusion}
We proposed a unified Expectation-Maximization framework for robust 
covariance estimation in the Student-t subclass of SIRV models with 
missing data, leveraging conjugate inverse-gamma priors to obtain 
closed-form updates for both the E-step and M-step. Validation on 
synthetic data and real Sentinel-1 interferograms demonstrates that 
EM-SIRV consistently outperforms mean imputation, PCA-based methods, 
and standard Gaussian EM across all scenarios, including a 
coherence-based MNAR self-masking regime that formally violates the 
MAR assumption, confirming the practical robustness of the framework 
beyond its theoretical assumptions. Future work will focus on 
explicitly integrating the missingness mechanism into the likelihood, 
deriving identifiability conditions for joint estimation of the 
covariance and missingness parameters, and extending the framework to 
other SIRV distributions beyond the Student-t model.
\FloatBarrier
\clearpage
\bibliographystyle{IEEEtran}
\bibliography{SIRV}

@article{dempster1977maximum,
  title={Maximum likelihood from incomplete data via the EM algorithm},
  author={Dempster, Arthur P and Laird, Nan M and Rubin, Donald B},
  journal={Journal of the royal statistical society: series B (methodological)},
  volume={39},
  number={1},
  pages={1--22},
  year={1977},
  publisher={Wiley Online Library}
}

@article{chitour2008exact,
  title={Exact maximum likelihood estimates for SIRV covariance matrix: Existence and algorithm analysis},
  author={Chitour, Yacine and Pascal, Fr{\'e}d{\'e}ric},
  journal={IEEE Transactions on signal processing},
  volume={56},
  number={10},
  pages={4563--4573},
  year={2008},
  publisher={IEEE}
}

@book{delmas2024elliptically,
  title={Elliptically symmetric distributions in signal processing and machine learning},
  author={Delmas, Jean-Pierre and El Korso, Mohammed Nabil and Fortunati, Stefano and Pascal, Fr{\'e}d{\'e}ric},
  volume={1},
  year={2024},
  publisher={Springer}
}

@article{conte2002recursive,
  title={Recursive estimation of the covariance matrix of a compound-Gaussian process and its application to adaptive CFAR detection},
  author={Conte, Ernesto and De Maio, Antonio and Ricci, Giuseppe},
  journal={IEEE Transactions on signal processing},
  volume={50},
  number={8},
  pages={1908--1915},
  year={2002},
  publisher={IEEE}
}

@article{kondrashov2006spatio,
  title={Spatio-temporal filling of missing points in geophysical data sets},
  author={Kondrashov, Dmitri and Ghil, Michael},
  journal={Nonlinear Processes in Geophysics},
  volume={13},
  number={2},
  pages={151--159},
  year={2006},
  publisher={Copernicus Publications G{\"o}ttingen, Germany}
}

@article{gerber2018predicting,
  title={Predicting missing values in spatio-temporal remote sensing data},
  author={Gerber, Florian and de Jong, Rogier and Schaepman, Michael E and Schaepman-Strub, Gabriela and Furrer, Reinhard},
  journal={IEEE Transactions on Geoscience and Remote Sensing},
  volume={56},
  number={5},
  pages={2841--2853},
  year={2018},
  publisher={IEEE}
}

@article{beckers2003eof,
  title={EOF calculations and data filling from incomplete oceanographic datasets},
  author={Beckers, Jean-Marie and Rixen, Michel},
  journal={Journal of Atmospheric and oceanic technology},
  volume={20},
  number={12},
  pages={1839--1856},
  year={2003}
}

@article{hippert2020eof,
  title={EM-EOF: Gap-filling in incomplete SAR displacement time series},
  author={Hippert-Ferrer, Alexandre and Yan, Yajing and Bolon, Philippe},
  journal={IEEE Transactions on Geoscience and Remote Sensing},
  volume={59},
  number={7},
  pages={5794--5811},
  year={2020},
  publisher={IEEE}
}

@article{chen2018robust,
  title={Robust covariance and scatter matrix estimation under Huber’s contamination model},
  author={Chen, Mengjie and Gao, Chao and Ren, Zhao},
  journal={The Annals of Statistics},
  volume={46},
  number={5},
  pages={1932--1960},
  year={2018},
  publisher={JSTOR}
}

@article{josse2012handling,
  title={Handling missing values in exploratory multivariate data analysis methods},
  author={Josse, Julie and Husson, Fran{\c{c}}ois},
  journal={Journal de la soci{\'e}t{\'e} fran{\c{c}}aise de statistique},
  volume={153},
  number={2},
  pages={79--99},
  year={2012}
}

@article{ollila2012complex,
  title={Complex elliptically symmetric distributions: Survey, new results and applications},
  author={Ollila, Esa and Tyler, David E and Koivunen, Visa and Poor, H Vincent},
  journal={IEEE Transactions on signal processing},
  volume={60},
  number={11},
  pages={5597--5625},
  year={2012},
  publisher={IEEE}
}

@misc{esa_snap,
  author = {{European Space Agency (ESA)}},
  title = {{SNAP}},
  howpublished = {\url{https://step.esa.int/main/toolboxes/snap}},
}

@article{pascal2008covariance,
  author  = {Pascal, F. and Chitour, Y. and Ovarlez, J.-P. and Forster, P. and Larzabal, P.},
  title   = {Covariance Structure Maximum-Likelihood Estimates in Compound {G}aussian Noise},
  journal = {IEEE Trans. Signal Process.},
  volume  = {56},
  number  = {1},
  pages   = {34--48},
  year    = {2008}
}

@article{tyler1987distribution,
  author  = {Tyler, D. E.},
  title   = {A Distribution-Free {M}-Estimator of Multivariate Scatter},
  journal = {Ann. Statist.},
  volume  = {15},
  pages   = {234--251},
  year    = {1987}
}

@article{hippert2022robust,
  author  = {Hippert-Ferrer, A. and {El Korso}, M. N. and Breloy, A. and Ginolhac, G.},
  title   = {Robust Low-Rank Covariance Matrix Estimation with a General Pattern of Missing Values},
  journal = {Signal Processing},
  volume  = {198},
  pages   = {108460},
  year    = {2022}
}

@book{little2002statistical,
  author    = {Little, R. J. A. and Rubin, D. B.},
  title     = {Statistical Analysis with Missing Data},
  edition   = {2nd},
  publisher = {Wiley},
  address   = {New York},
  year      = {2002}
}

@article{frahm2010generalization,
  author  = {Frahm, G. and Jaekel, U.},
  title   = {A Generalization of {T}yler's {M}-Estimators to the Case of Incomplete Data},
  journal = {Comput. Statist. Data Anal.},
  volume  = {54},
  pages   = {374--393},
  year    = {2010}
}

@article{sun2014regularized,
  title={Regularized Tyler's scatter estimator: Existence, uniqueness, and algorithms},
  author={Sun, Ying and Babu, Prabhu and Palomar, Daniel P},
  journal={IEEE Transactions on Signal Processing},
  volume={62},
  number={19},
  pages={5143--5156},
  year={2014},
  publisher={IEEE}
}

@book{kotz2004multivariate,
  title={Multivariate t-distributions and their applications},
  author={Kotz, Samuel and Nadarajah, Saralees},
  year={2004},
  publisher={Cambridge university press}
}

@article{zebker1992decorrelation,
  title={Decorrelation in interferometric radar echoes},
  author={Zebker, HOWARDA and Villasenor, John and others},
  journal={IEEE Transactions on geoscience and remote sensing},
  volume={30},
  number={5},
  pages={950--959},
  year={1992}
}

@article{bamler1998synthetic,
  title={Synthetic aperture radar interferometry},
  author={Bamler, Richard and Hartl, Philipp},
  journal={Inverse problems},
  volume={14},
  number={4},
  pages={R1--R54},
  year={1998}
}

@article{meriaux2020matched,
  title={Matched and mismatched estimation of kronecker product of linearly structured scatter matrices under elliptical distributions},
  author={Meriaux, Bruno and Ren, Chengfang and Breloy, Arnaud and El Korso, Mohammed Nabil and Forster, Philippe},
  journal={IEEE Transactions on Signal Processing},
  volume={69},
  pages={603--616},
  year={2020},
  publisher={IEEE}
}

@article{abdallah2019detection,
  title={Detection methods based on structured covariance matrices for multivariate SAR images processing},
  author={Abdallah, R Ben and Mian, Ammar and Breloy, Arnaud and Taylor, Abiga{\"e}l and El Korso, Mohammed Nabil and Lautru, David},
  journal={IEEE Geoscience and Remote Sensing Letters},
  volume={16},
  number={7},
  pages={1160--1164},
  year={2019},
  publisher={IEEE}
}

@article{hippert2025missing,
  title={Missing Data in Signal Processing and Machine Learning: Models, Methods and Modern Approaches},
  author={Hippert-Ferrer, Alexandre and Sportisse, Aude and Javaheri, Amirhossein and Korso, Mohammed Nabil El and Palomar, Daniel P},
  journal={arXiv preprint arXiv:2506.01696},
  year={2025}
}

@article{cherifi2025robust,
  title={Robust inference with incompleteness for logistic regression model},
  author={Cherifi, M and El Korso, Mohammed Nabil and Fortunati, Stefano and Mesloub, Ammar and Ferro-Famil, Laurent},
  journal={Signal Processing},
  volume={236},
  pages={110027},
  year={2025},
  publisher={Elsevier}
}

\end{document}